\documentclass[12pt]{article}
\usepackage{amssymb,amsmath}
\usepackage[utf8]{inputenc}
\usepackage{amsfonts}
\usepackage[pdftex]{color,graphicx}
\numberwithin{equation}{section}
\def\BigRoman{\uppercase\expandafter{\romannumeral\number\count 255 }}
\def\Romannumeral{\afterassignment\BigRoman\count255=}
\begin{document}

\title{\bf Toward {practical/realistic} Randall-Sundrum Brane world scenario.
} 
\author {\bf Uicheol Jang$\footnote{Astronomy \&Space science and Geology, Chungnam Natinal Univ.}$ \  {\bf Hongsu Kim$\footnote{Center for Theoretical Astronomy, KASI}$} }

\date{2020} 
\maketitle
\begin{abstract}
The Randall-Sundrum brane world scenario which has been put forward back in 1998 was indeed a remarkable paradigm change from the Kaluza-Klein theory which has been a long standing idea for the unification of gravitation and E\&M.\\
The real breakthrough was replacing the compact fiber manifold $S^1$ by an open infinite $R^1$. 
This rather naive looking new challenge was successful because the graviton gets trapped by the deep volcano type potential which is surrounding the brane world that is supposed to be our universe in which we live . 4-dimensional gravitation is still secured even after unfolding $S^1$ to $R^1$. Such a brand new paradigm is indeed very fresh and remarkable enough to excite the whole theoretical physics community. Their new setup, however, is not completely without any shortcomings as they assumed their brane world to be Poincare-invariant.
This is indeed a serous drawback as it implies that  the brane world which supposed to be our universe is totally cold and empty. Obviously our universe is filled with various type of structures and even it is expanding. Therefore from a critical standpoint to elevate this original set-up to a more realistic and practical set up here in this work we replace the Poincare-invariance with the Ricci flat set up for the brane world. To be more concrete the Ricci flat set up allows for some space time structure and universe evolution behavior such as black hole and Bianchi type 9 universe models, respectively, which are the solutions of vacuum Einstein equations.\\
Remarkably enough, even upon this rich elaboration, the Randall-Sundrum brane world scenario is mostly preserved and still secure as we realized that the graviton trap is still safely guaranteed. We would like to empathize here that such necessary but rich elaboration doesn't cost a fortune as the Ricci flatness condition effectively accommodates the realistic structure of our universe on the one hand, while still sustaining the 4-dimensional graviton trap on the other. 
 \end{abstract}

\section{Introduction}
The Paradigm change by Randall-Sundrum brane world scenario in higher-dimensional gravitation can be summarized the following .\\
\subsection{Kaluza-Klein Unification.}
The extra $5^{\textrm{th}}$ dimension(space-like) has to be compact($S^1$)-(Kaluza) and infinitesimal-(Klein) in order for it  not  to be observable disturbing our 4-dim. world.\\
\subsection{Randall-Sundrum brane world scenario.}
$\rightarrow 5^{\textrm{th}}$ dimension has to be neither compact nor Planck-sized; indeed, it may be non-compact($R^1$) and infinite\\
as the 4-dim. gravity(graviton) gets localized(trapped) in the 4-dim. time-like Lorentzian submanifold(i.e.,"brane world")-(L.Randall\&R.Sundrum)!\\
the 4-dim. gravity(graviton) gets localized(trapped) on the Poincare-invariant, 4-dim. time-like Lorentzian  sub-manifold(i.e.,"brane world")-(L.randall\&L.Sundrum); \\
thus the original version of RS brane world is a cold, empty universe?\\
\subsection{Elaboration of Randall-Sundrum brane world scenario.}
Our elaboration on this Randall-Sundrum" brane world scenario-(ongoing work, H.Kim)\\
$\rightarrow$ the 4 dim. gravity(graviton) gets localized(trapped) on the Ricci-flat, 4dim. time-like Lorentzian submanifold(i.e."brane world");\\
the RS brane world could be promoted to our real expanding world having structure(e,g,. black holes) as we now can upload any geometry on it as long as it is a solution of the 4-dim. vacuum Einstein equation and hence Ricci-flat ($R_{\mu\nu} = 0$)!\\
(a cosmological implication) a homogeneous, anisotropic cosmology(i.e., Kasner universe model with the Hawking-Penrose singularity - the "B-K-L" initial time singularity) can live on the brane world-just like our conventional wisdom for the universe creation!!\\
To summarize, Althogh the higher dimensional gravitation paradigm change from the long standing Kaluzs-Klein's to Randall-Sundrum's was indeed a remarkable breakthrough. 
Randall-Sundrum's scenario itself it NOT without serious flaws and hence the present weak of H.Kim attempts to correct them for well defined, final revision of Randall-Sundrum scenario. 

\section{4-dim. gravity(graviton) localization(trap) on the Ricci-flat, time-like Lorentzian sub-manifold(i.e.,"brane world")}

\begin{align*}
S=&S_{\textrm{bulk}} + S_{\textrm{brane}}  + S_{\textrm{brane'}} \\
(\textrm{with})& \\
S&=\int d^4x \int_{-\pi}^{+\pi}d\phi\sqrt{G}[2M^3R-\Lambda]\\&\left<\begin{array}{ll} S_{\textrm{brane}} = \int d^4x\sqrt{G_{\textrm{brane}}} [L_{\textrm{brane}}-V_{\textrm{brane}}],\\
S_{\textrm{brane`}} = \int d^4x\sqrt{G_{\textrm{brane`}}} [L_{\textrm{brane`}}-V_{\textrm{brane`}}]
\end{array}\right.\\
\textrm{where} &\\
&g_{\mu\nu}^{\textrm{brane}}(x^\mu) = G_{\mu\nu}(x^\mu, \phi=0)\\
&g_{\mu\nu}^{\textrm{brane`}}(x^\mu) = G_{\mu\nu}(x^\mu, \phi=\pi)\\
\end{align*}
with $G_{AB}, A,B=\mu,\phi$ is the 5-dim. bulk metric\\
Next, the 5-dim Einstein equation that one upon extremizing this action for our system reads;
\begin{align*}
\sqrt{G}[R_{AB}-\frac{1}{2}RG_{AB}]=\\
\frac{-1}{4M^3}[\sqrt{G}\Lambda G_{AB}&+V_{\textrm{brane}}\sqrt{g}g_{\mu\nu}^{\textrm{brane}}\delta_A^\mu\delta_B^\nu\delta(\phi)\\
&+V_{\textrm{brane'}}\sqrt{g}g_{\mu\nu}^{\textrm{brane'}}\delta_A^\mu\delta_B^\nu\delta(\phi-\pi)]
\end{align*}
when \\

\[\sigma(\phi)=kr_c|\phi|=k|y|\]

we now look for a solution corresponding to a "non-factorizable" metric with "warp-factor"
\begin{align*}
ds^2&=e^{-2\sigma(\phi)}\gamma_{\mu\nu}(x)dx^\mu dx^\nu+r_c^2d\phi^2\\
&=e^{-2\sigma(y)}\gamma_{\mu\nu}(x)dx^\mu dx^\nu+dy^2\\
\end{align*}
where $y\equiv r_c\phi$ is the 5th dimension coordinate transverse to  the brane. 
\begin{equation*}
V_{\textrm{brane}}=-V_{\textrm{brane`}}=24M^3k, \Lambda=-24M^3k^2
\end{equation*}
when\\
\[ \sigma(\phi)=kr_c|\phi|=k|y|\]\\
Finally, the bulk metric solution is given by 
\begin{align*}
ds^2&=e^{-2\sigma(\phi)}\gamma_{\mu\nu}(x)dx^\mu dx^\nu+r_c^2d\phi^2\\
&=e^{-2\sigma(y)}\gamma_{\mu\nu}(x)dx^\mu dx^\nu+dy^2\\
\end{align*}
(where) $\hat{R}_{\mu\nu}(r)=0 $ and we assume that $K<M$(with $M$ being the fundamental Planck scale in 5-dim.) so that the bulk curvature ($R\sim\Lambda=-24M^3k^2$) is "small" compared to higher
dimensional Planck scale($\sim M^5$) and we trust our solution as a classical one.
\[ G_{\mu\nu} = e^{-2k(y)}\gamma_{\mu\nu}(x)+h_{\mu\nu}(x,y)\]
Let $h(x,y)=e^{ipx}\psi(y)$ (namely general fluctuations can be written as superpositions of modes) and use $P^2=P^\alpha P_\alpha=-(P^0)^2+\vec{P}^2=-m^2, $then
\[ [-\frac{m^2}{2}e^{2k|y|}-2\frac{1}{2}\partial_y^2+2k^2-2k\delta(y)]\psi(y)=0\]
$\Rightarrow$\[ [-\frac{1}{2}\frac{d^2}{dz^2}+V(z)]\hat{\psi}(z)=\frac{m^2}{2}(z)\]
with
\[ V(z)=\{\frac{15k^2}{8(k|z|+1)^2}-\frac{3k}{2}\delta(z)\}\]

\begin{figure*}
\centering
\includegraphics[ width=1 \textwidth]{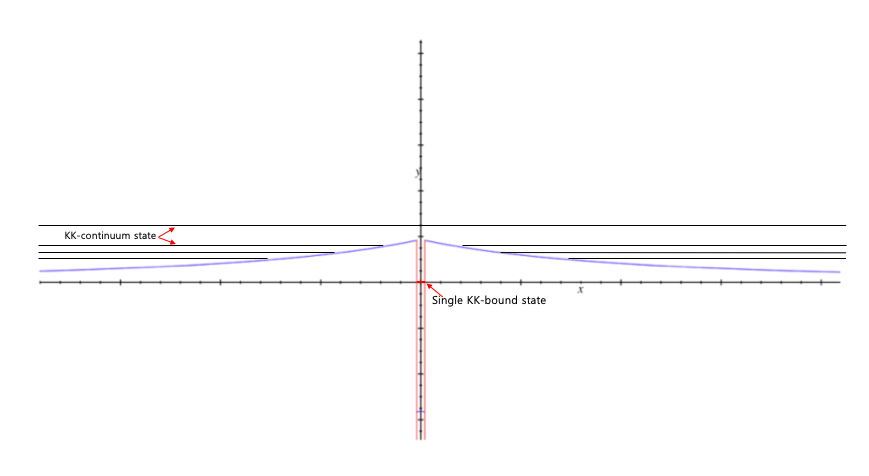}
\caption{Valcano-type potential for KK-excited states}
\end{figure*} 
\newpage	
\section{Application}
{\large {\bf BKL oscillatory singularity} on the Ricci flat brane}\\
Belinskii-Khalatnikov-Lifshitz(BKL) oscillatory singularity representing anisotropic cosmology approaching initial time singularity at the early universe\\
Initial time singularity(singularity theorem):\\
"-if Einstein equation holds, and \\
-weak/strong energy condition is satisfied, but without\\
-the emergence of closed timelike curves, then as we move backward in time, we approach the initial time singluarity where the proper volume element of a spacelike hyper-surface vanishes!"\\

Hawking and Penrose(1965~1970) $\rightarrow$ they mathematically prove the singularity theorem in terms of topology employing the Raychaudhuri's equation.\\
B-K-L(1968)$\rightarrow$ they physically exhibited the singularity theorem in terms of geometry by explicitly solving the vacuum Einstein equation employing the Bianchi type-IX cosmology model 
\\
more on Belinskii-Khalatnikov-Lifshitz(BKL) oscillatory singularity\\
B-K-L version of initial time singularity where the proper volume element of a spacelike hypersurface, which goes like $\sim$t, vanishes!\\

B-K-L(1968) $\Rightarrow$ they discovered that if we model the early universe by employing the Bianchi type-IX cosmology metric, then as the initial time singularity is approached(t$\rightarrow$ 0.), one inevitably encounters successive series of Kasner regimes during each of which the distances along two of the three spatial axes oscillate, while they fall/drop off monotonically along the third and the spatial volume element drops off approximately as$\sim$t. And in going from one Kasner regime to the next, the direction along which there is a monotonic drop off of distances shifts from one spatial direction to another. And as the initial time singularity is approached, the order of these shifts takes on the character of a random process, namely,a "chaotic map" \\
\\
at early stages of the universe, we employ the Bianchi type-IX(mixmaster) cosmology model representing a homogeneous but not necessarily isotropic, spatially-closed geometry <homogeneous but anisotropic universe model>
\begin{align*}
ds^2=&-N^2(t)dt^2+e^{2\alpha(t)}(e^{2\beta(t))}_{ij}\sigma^i\otimes\sigma^j\\
=&-N^2dt^2+e^{2\alpha}[e^{2\beta_1}\sigma_1^2+e^{2\beta_2}\sigma_2^2+e^{2\beta_3}\sigma_3^2]\\
=&-N^2dt^2+[A^2(t)\sigma_1^2+B^2(t)\sigma_2^2+C^2(t)\sigma_3^2]
\end{align*}
where $N(t)$ is the lapse function, $e^{\alpha(t)}$ is the scale factor of the spatial section and the symmetric, traceless matrix $\beta_{ij}(t)$ satisfying $\det(e^{2\beta)}=1$ (or$ tr(\beta)=0$) parametrizes the anisotropy. This matrix $\beta_{ij}$ may be chosen to be and it is convenient to define them explicitly by 
\begin{align*}
\beta_{11}\equiv &\beta_1=\beta_++\sqrt{3}\beta_-,\\
\beta_{22}\equiv &\beta_2=\beta_+-\sqrt{3}\beta_-,\\
\beta_{33}\equiv &\beta_3=-2\beta_+,\\
\end{align*}
The non-holonomic basis\{$\sigma^i\} (i=1,2,3)$ from a basis Maurer-Cartan structure equation
\[ d\sigma^i -\frac{1}{2}\epsilon^{ijk}\sigma^i\Lambda\sigma^k\]

To summarize, the 4-Einstein equations can be put to the forms;
\begin{align}
2\ddot a=(B^2-C^2)^2-A^4\\
2\ddot b=(C^2-A^2)^2-B^4\\
2\ddot c=(A^2-B^2)^2-C^4
\end{align}

which, upon being added up, gives
\begin{equation}
 2(\ddot a+\ddot b+\ddot c)=[A^4+B^4+C^4-2(A^2B^2+B^2C^2+C^2A^2)]
 \end{equation}
 (3.4)=(3.1)+(3.2)+(3.3) and
 \begin{equation}
 \dot{ab}+\dot{bc}+\dot{ba}=\frac{1}{4}[A^4+B^4+C^4-2(A^2B^2+B^2C^2+C^2A^2)]
 \end{equation}
 
\begin{align*}
A(t)&=e^{a(t)} \equiv e^{(\alpha+\beta_1)}\\
B(t)&=e^{b(t)} \equiv e^{(\alpha+\beta_2)}\\
C(t)&=e^{c(t)} \equiv e^{(\alpha+\beta_3)}.
\end{align*}

Therefore, we first assume that at our starting point corresponding to some early universe stage, the(r.h.s) of egs, (3.1)$\sim$(3.4) were negligibly small and hence the metric functions essentially exhibit the behaviors
\[A\simeq\tau^{P_l},B\simeq\tau^{P_m},C\simeq\tau^{P_n},\]
where we introduce another time parameter $d\tau=Ndt=(ABC)dt$ and ($P_l,P_m,P_n$) are numbers satisfying the relations
\[P_l+P_m+P_n = 1= P_l^2+P_m^2+P_n^2\]
Thus our starting  point in some early universe may be regarded as a Kasner-type epoch.\\
Point on, we shall retain the notation ($P_1,P_2,P_3$) for he triple of numbers arranged in the order $P_1<P_2<P_3$ and taking on values
\[ -\frac{1}{3} \leq P_1\le0 \le P_2\le \frac{2}{3} \le P_3\le 1\]

Note that these numbers can be written in parametric form as
\begin{equation}
P_1(u)=\frac{-u}{1+u+u^2},P_2(u)=\frac{u+1}{1+u+u^2} ,P_3(u)=\frac{u(u+1)}{1+u+u^2}
 \end{equation}
 then all the different value of $P_1,P_2,P_3$ (preserving the assumed order) are obtained if the parameter "$u$"runs in the range $u\ge1$, The values for $u\le1$ are reduced to this same region as fallows
 \begin{equation}
 P_1(\frac{1}{u})=P_1(u), P_2(\frac{1}{u})=P_3(u), P_3(\frac{1}{u})=P_2(u)
 \end{equation}

\newpage

{\large\bf[Concluding Remarks]}\\

Our claim in the present work can be summarized as follows.
 The 4-dim. gravity(graviton) gets localized (trapped) on the Poincare- invariant, 4-dim. time-like Lorentzian submanifold(i.e brane world")-(L. Randall \& R. Sundrum)\\ 
\\
  Can be generalized (relaxed) to a more generic case where \\ \\
  The 4-dim. gravity (graviton) gets localized (trapped) on the Ricci-flat, 4- dim. time-like Lorentzian submanifold (i.e., brane world") \\ \\
  A result \\ \\
  
  The original version of RS brane world is a cold, empty universe ?[unphysical] \\ \\
  
  The RS brane world could now be promoted to a much more our real expanding realistic/physical  universe having structures (e.g., black holes/GW)

\end{document}